\documentclass[%
 reprint,
 amsmath,amssymb,
 aps,
pra,
]{revtex4-1}

\usepackage{graphicx}% Include figure files
\usepackage{dcolumn}% Align table columns on decimal point
\usepackage{bm}% bold math
\begin{document}

\preprint{APS/123-QED}

%\title{Inhibited electron spin diffusion in n-doped CdTe revealed by spin noise spectroscopy }% Force line breaks with \\
%\title{Transition from long to short electron spin correlation time revealed by spin noise spectroscopy in n-CdTe }%
\title{Spatiotemporal electronic spin fluctuations in random nuclear fields in n-CdTe}

\author{S. Cronenberger, C. Abbas and D. Scalbert}%
 \affiliation{Laboratoire Charles Coulomb (L2C), UMR 5221 CNRS-Universit\'{e} de Montpellier, Montpellier, FR-34095, France}%Lines break automatically or can be forced with \\
\author{H. Boukari}
\affiliation{Univ. Grenoble Alpes, F-38000 Grenoble, France}
\affiliation{CNRS, Institut NEEL, F-38000 Grenoble, France}
%\author{D. Smirnov}
%\affiliation{Ioffe Institute, 194021, St.-Petersburg, Russia}
\date{\today}% It is always \today, today,

\begin{abstract}
We report on the dynamics of electron spins in n-doped CdTe layers that differs significantly from the expected response derived from the studies dedicated to electron spin relaxation in n-GaAs. At zero magnetic field, the electron spin noise spectra exhibit a two-peak structure - a zero-frequency line and a satellite - that we attribute to the electron spin precession in a frozen random nuclear spin distribution. This implies a surprisingly long electron spin correlation time whatever the doping level, even above the Mott transition. Using spatiotemporal spin noise spectroscopy, we demonstrate that the observation of a satellite in the spin noise spectra and a fast spin diffusion are mutually exclusive. This is consistent with a shortening of the electron spin correlation time due to hopping between donors. We interpret our data via a model assuming that the low temperature spin relaxation is due to hopping between donors in presence of hyperfine and anisotropic exchange interactions. Most of our results can be interpreted in this framework. First, a transition from inhomogeneous to homogeneous broadening of the spin noise peaks and the disappearance of the satellite are observed when the hopping rate becomes larger than the Larmor period induced by the local nuclear fields. In the regime of homogeneous broadening the ratio between the spin diffusion constant and the spin relaxation rate has a value in good agreement with the Dresselhaus constant. In the regime of inhomogeneous broadening, most of the samples exhibit a broadening consistent with the distribution of local nuclear fields. We obtain a new estimate of the hyperfine constants in CdTe and a value of 0.10 Tesla for the maximum nuclear field. Finally, our study also reveals a puzzle as our samples behave as if the active donor concentration was reduced by several orders of magnitudes with respect to the nominal values.

%\begin{description}

%\verb+\pacs{#1}+ command.
%\end{description}
\end{abstract}

\maketitle

%\tableofcontents

\section{Introduction}
A complete understanding of spin relaxation mechanisms in semi-conductors is mandatory for the development of spintronics applications using these materials. Although this subject has been investigated for several decades it is still an active field of research. First historical works identified the relevant spin relaxation mechanisms, and their relative efficiencies mainly in bulk III-V compounds \cite{Meier1984}. More recent studies focused on structures of reduced dimensionality, which offer a plethora of possibilities for the control of the spin relaxation \cite{Dyakonov2017,Dyakonov1986b, Ohno1999,Sandhu2001,Balocchi2011}. One could now consider that the electron spin relaxation is generally well understood in semiconductors although most of the studies have been devoted to GaAs  \cite{Kikkawa1998,Dzhioev2002,Kavokin2008,Lonnemann2017,Harley2017}.  Typically at low temperatures and below the metal insulator transition (MIT) two spin relaxation mechanisms compete in this compound. The spin relaxes either because of its random evolution in the fluctuating hyperfine fields \cite{Merkulov2002}, or because it rotates during hopping between donors in presence of anisotropic exchange interaction \cite{Kavokin2001}. Above the MIT the Dyakonov-Perel mechanism becomes the dominant relaxation mechanism, and explains very well the measured spin relaxation times, including in the metallic impurity regime if one takes into account the change of the scattering factor with the doping density \cite{Lonnemann2017}, and eventually the electron-electron scattering which comes into play at higher temperature \cite{Leyland2007}.

This understanding is far from being reached in the case of CdTe where studies are scarce  \cite{Sprinzl2010,Linpeng2016,Garcia-Arellano2019}. However, because GaAs and CdTe possess similar band structures this understanding should be in principle directly transposable from one to the other.

We present here results which can be hardly reconciled with this expectation. While the dominating spin relaxation channel depends critically on the density of donors in GaAs, we found that the spin relaxation in CdTe is dominated by the hyperfine interaction, at low power and low temperatures, and for a wide range of doping levels. In addition, we observe, in most of the studied samples,a two-peak structure in the zero magnetic field spin noise spectra, which is imputable to the evolution of the electron spin in a random local nuclear field. The appearance of a satellite in the spin noise spectra requires an electron spin correlation time longer than the Larmor precession period in the local nuclear field. In GaAs this condition could be reached only in a very pure sample containing a very low density of donors \cite{Berski2015a}.

A deeper investigation of the electron spin dynamics in CdTe, using the technique of spatiotemporal spin noise
spectroscopy (SNS), which we have recently developed \cite{Cronenberger2019}, is also reported here. This approach allows us to measure simultaneously the electron spin relaxation rate, and the spin diffusion constant. Moreover we estimate the correlation time of the electron spin on a given donor site from the shape of the two-peak structure at zero-field.

We find that the two-peak structure at zero magnetic field shows up only when the spin diffusion is slow enough. This corresponds to a transition from the short to the long electron spin correlation time $\tau_c$, which is clearly revealed in the spin noise spectra. The long correlation times regime at $\omega\tau_c>1$, where $\omega$ is the angular precession frequency in the local nuclear fields, is characterized by three main features:
  \begin{enumerate}
   \item The two-peak structure is observed in zero-field.
   \item In transverse magnetic field larger than the local nuclear fields the spin noise spectrum exhibits a gaussian lineshape, with a FWHM of about 48~MHz.
   \item The spin diffusion constant is smaller than $0.4\text{ cm}^2/\text{s}$.
 \end{enumerate}
On the contrary in the short correlation time regime, the two-peak structure disappears, the line becomes lorentzian, and the spin diffusion is fast.

The paper is organized as follows. In Sec.~\ref{Sec:samples&method} we present the studied samples and give some details on the experimental technique. In Sec.~\ref{Sec:Obs_satellites} we demonstrate the existence of a two-peak structure in the zero-field noise spectra, one line at zero-frequency, and one satellite (two in case of heterodyne experiments). We study the temperature and power dependencies of the two-peak structure, and show that it can be interpreted in terms of electron spin precession in the local nuclear field at each occupied donor site. Section~\ref{Sec:diffusion} is devoted to the determination of the spin diffusion constant. In Sec.~\ref{Sec:interpretation} we tentatively interpret our data considering that spin relaxation is caused by hyperfine and spin-orbit interaction in presence of hopping between donors. In Sec.~\ref{Sec:longitudinal} we present our results concerning the longitudinal spin relaxation time. Conclusions and perspectives to this work are given in Section~VII.

\section{Samples and experimental methods}\label{Sec:samples&method}

\subsection{Samples}

\begin{table}
  \centering
  \begin{tabular}{|c|c|c|c|c|}
  \hline
  sample & $n_D$  & thickness & substrate & satellite\\
  & (cm$^{-3}$) & ($\mu$m) & & \\
  \hline
  M1409 & $6\times 10^{18}$ & 0.58 & Cd$_{0.96}$Zn$_{0.04}$Te & yes (B) \\
  M896 & $6\times 10^{16}$ & 0.6 & Cd$_{0.96}$Zn$_{0.04}$Te & yes (B) \\
  M1371 & $6\times 10^{15}$ & 0.52 & Cd$_{0.96}$Zn$_{0.04}$Te & yes (B) \\
  M1559 & $3\times 10^{16}$ & 2.5 & Cd$_{0.96}$Zn$_{0.04}$Te & yes (B+F) \\
  M3405 & $3\times 10^{17}$ & 13.3 & CdTe & yes (B) \\
  M3406 & $\sim 10^{16}$ & 13.3 & CdTe & no (2.6 mW) \\
  M3407 & $\sim 10^{15}$ & 13.3 & CdTe & yes (B) \\
  M3408 & $2\times 10^{17}$ & 0.5 & CdTe & yes (B) \\
  M3416 & $\sim 10^{17}$ & 13.5 & Cd$_{0.96}$Zn$_{0.04}$Te & no (B+F) \\
  M3459 & $10^{17}$ & 10 & Cd$_{0.96}$Zn$_{0.04}$Te & no (F) \\
  M3460 & $2\times 10^{16}$ & 10 & Cd$_{0.96}$Zn$_{0.04}$Te & yes (F) \\
  \hline
\end{tabular}
  \caption{List of the studied samples with corresponding electron density $n_d$, thickness of the Al-doped CdTe layers and substrate. The last column indicates if the satellite has been detected (yes) or not (no) in the following configurations: B for backscattering and F for forward scattering.}\label{table:sample}
\end{table}

Table~\ref{table:sample} summarizes the main characteristics of the studied samples. All samples are Aluminium doped CdTe layers grown by molecular beam epitaxy on either CdTe(001) or Cd$_{0.96}$Zn$_{0.04}$Te(001) substrates.

 The four first samples are old samples in which we first observed the satellite. The substrate of sample M1559 has been mechanically polished to allow for measurements in forward scattering configuration. Note that in these samples, secondary ion mass spectrometry (SIMS) measurements revealed the presence of Zn in the doped CdTe layers (probably up to 2\% in sample M1371 revealed by a shift of the excitonic resonance of about 20~meV). In the case of sample M1559 clusters of Al were also detected.

The other samples were grown recently and specifically to study the appearance of the satellite in the SNS spectra. The donor density has been estimated from Hall measurements on the metallic sample M1409, which served as a reference for the determination of the concentration of the other samples that were characterized by SIMS. The rear side of the last three samples have been mechanically polished to allow for SNS measurements in forward scattering. The satellite has been detected in 8 samples out of 11 (see Table I) corresponding to an electron density varying across the Mott-insulator transition from $n_d\sim 10^{15}\text{ cm}^{-3}$ to $n_d\sim 6.10^{18}\text{ cm}^{-3}$.

%Let us outline that surface states may deplete partially the layers.
%The thickness of the depletion layer is given by $\ell_d=\sqrt{2\varepsilon U/e^2n_D}$, with $\varepsilon \sim 10$ the dielectric constant, $U$ the energy difference between the bottom of the conduction band and the Fermi energy at the surface, $e$ the electron charge, and $n_D$ the density of donors (supposed to be uncompensated). Taking $U\sim 1$~eV and for $n_D=2\times 10^{17}\text{ cm}^{-3}$ we find  $\ell_d \sim 100$~nm .

\subsection{Experimental methods}
Classically the optical detection of spin noise is explained in terms of Faraday noise induced by the spin fluctuations \cite{Crooker2004,Hubner2014}. But it can also be interpreted as a result of interference between the spin-flip Raman scattered light and the laser beam reflected or transmitted through the sample \cite{Gorbovitsky1983}. Therefore, the spin noise signal arises from homodyne mixing of the scattered light with the laser beam, which serves as a local oscillator (LO). We refer to this classical detection mode as self-homodyne detection. The probe beam can also be replaced by a LO for genuine homodyne or heterodyne detection, which offers more flexibility.

Here we use different detection configurations: self-homodyne (the standard method without LO), heterodyne \cite{Cronenberger2016}, and $q$-selective homodyne detection using spatiotemporal SNS, either in backscattering (B) or forward (F) scattering geometry \cite{Cronenberger2019}. The $q$-selective configuration allows to detect selectively spin fluctuations with a given wavevector, that gives directly access to spin diffusion. Here heterodyne detection is used as a convenient way to shift the noise spectra away from zero-frequency. This permits to measure the noise spectra at zero magnetic field without being bothered by the low frequency cut-off of our detector, and by low frequency noise.  For the heterodyne detection the local oscillator frequency $\nu_\text{LO}$ is shifted from the probe frequency $\nu_\text{P}$ by using either a fiber-coupled electro-optic intensity modulator (modulation frequency $\nu_\text{m}=500$~MHz). After modulation by the electro-optic modulator the beam passes though a Fabry-Perot to select one of the frequency-shifted laser lines.

\begin{figure}[ht]
  \centering
  \includegraphics[width=8 cm]{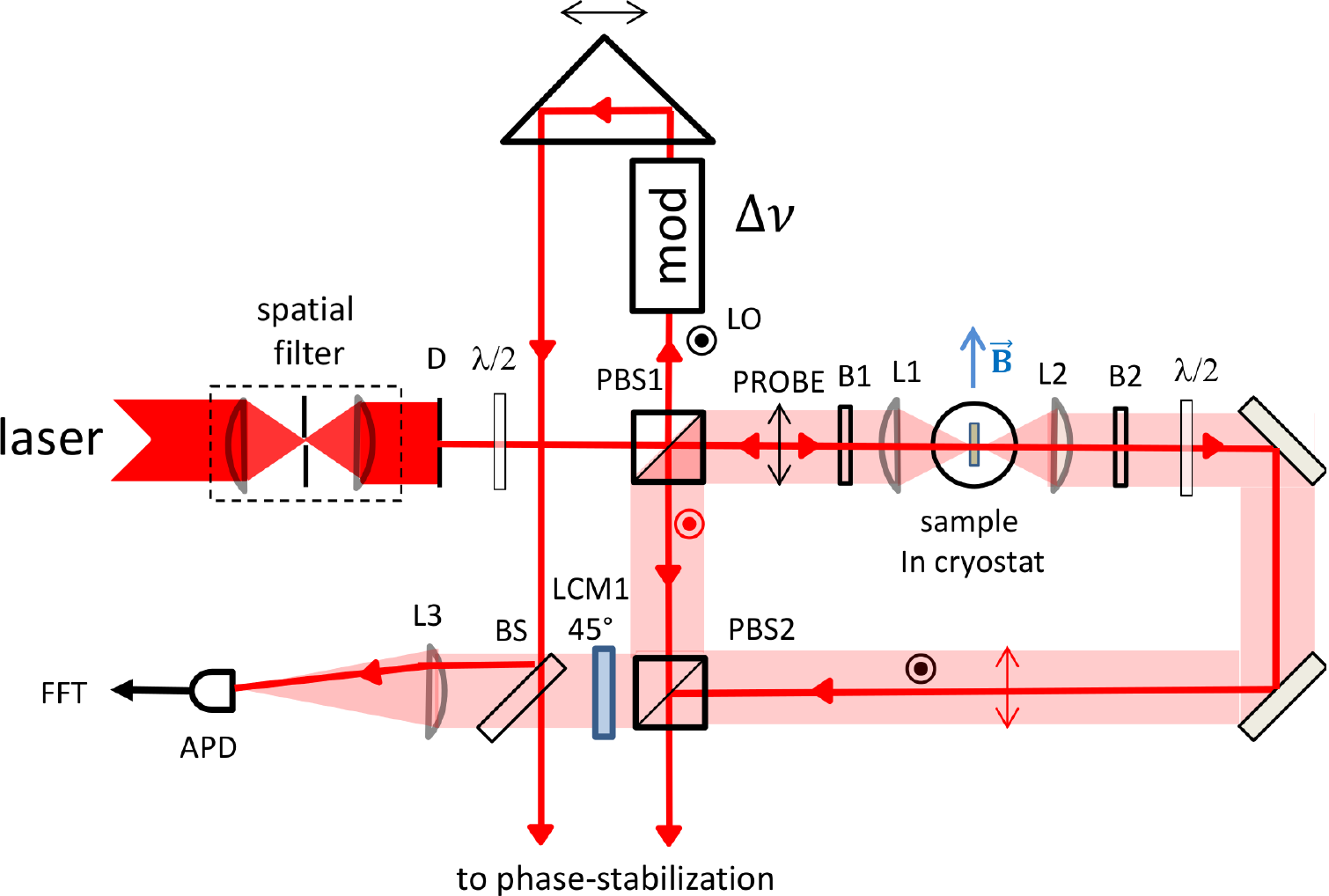}
  \caption{Schematics of the spin noise setup used for the different acquisition modes. A single mode and linearly polarized laser beam passes through a spatial filter and a diaphragm (D) to control the beam diameter. The beam is then split into a local oscillator (LO) and a probe beam by the polarizing beam splitter PBS1: polarizations are shown by black circle and arrow respectively. Their relative powers are controlled by the $\lambda/2$ waveplate placed before PBS1. B1 and B2 are Babinet compensators, LCM1 is a liquid crystal phase retarder, BS is a beam sampler which reflects 10\% of the incident light. The spin-flip Raman scattering light, which is cross-polarized with the probe (polarization shown in red), is represented in pink. Note that the polarizations of the transmitted probe and forward-scattered light are inter-changed by the $\lambda/2$ waveplate placed after B2. Most of the transmitted (or reflected) probe is cut by PBS2, while the scattered light propagates until the APD. LCM1 controls the polarization of the scattered light to allow homodyne mixing with LO or not depending on the applied voltage. The LO passes through an electro- or acousto-optic modulator, which shifts its frequency for heterodyne detection. The LO can be laterally shifted with a prism to control the wavevector of the detected spin fluctuation. The relative phase between probe and LO can be stabilized when necessary.}\label{setup}
\end{figure}

A scheme of the setup used for the different modes of spin noise detection is shown in Fig.~\ref{setup}. The polarizing beam splitters PBS1 and PBS2 allow the Raman light to reach the detector, while most of the reflected (or transmitted) laser light is suppressed. The Raman light is mixed on the detector with the LO for homodyne detection ($\nu_\text{LO}=\nu_\text{P}$) or heterodyne detection ($\nu_\text{LO}-\nu_\text{P}=\nu_\text{m}$), either at $q=0$ or $q\neq 0$. For self-homodyne detection the LO beam is blocked, and a part of the reflected (or transmitted) beam is allowed to reach the detector by slight detuning of one of the Babinet compensators.

The samples are in vacuum and glued on the cold finger of a He-flow cryostat. The temperatures indicated below are those measured at the heat-exchanger, and are presumably lower than the temperatures of the samples. In some experiments, a magnetic field $\overrightarrow{B}$ is applied in the plane of the n-doped CdTe layers.

Subtraction of the background noise is obtained either by taking the difference between noise spectra acquired at two different fields, or for homodyne detection by changing the polarization of the Raman light with a liquid crystal modulator (LCM1).

\section{Spin noise of electrons coupled to nuclei}\label{Sec:Obs_satellites}
\subsection{Observation of a satellite}\label{sec:satellite}

\begin{figure}[ht]
  \centering
  \includegraphics[width=8 cm]{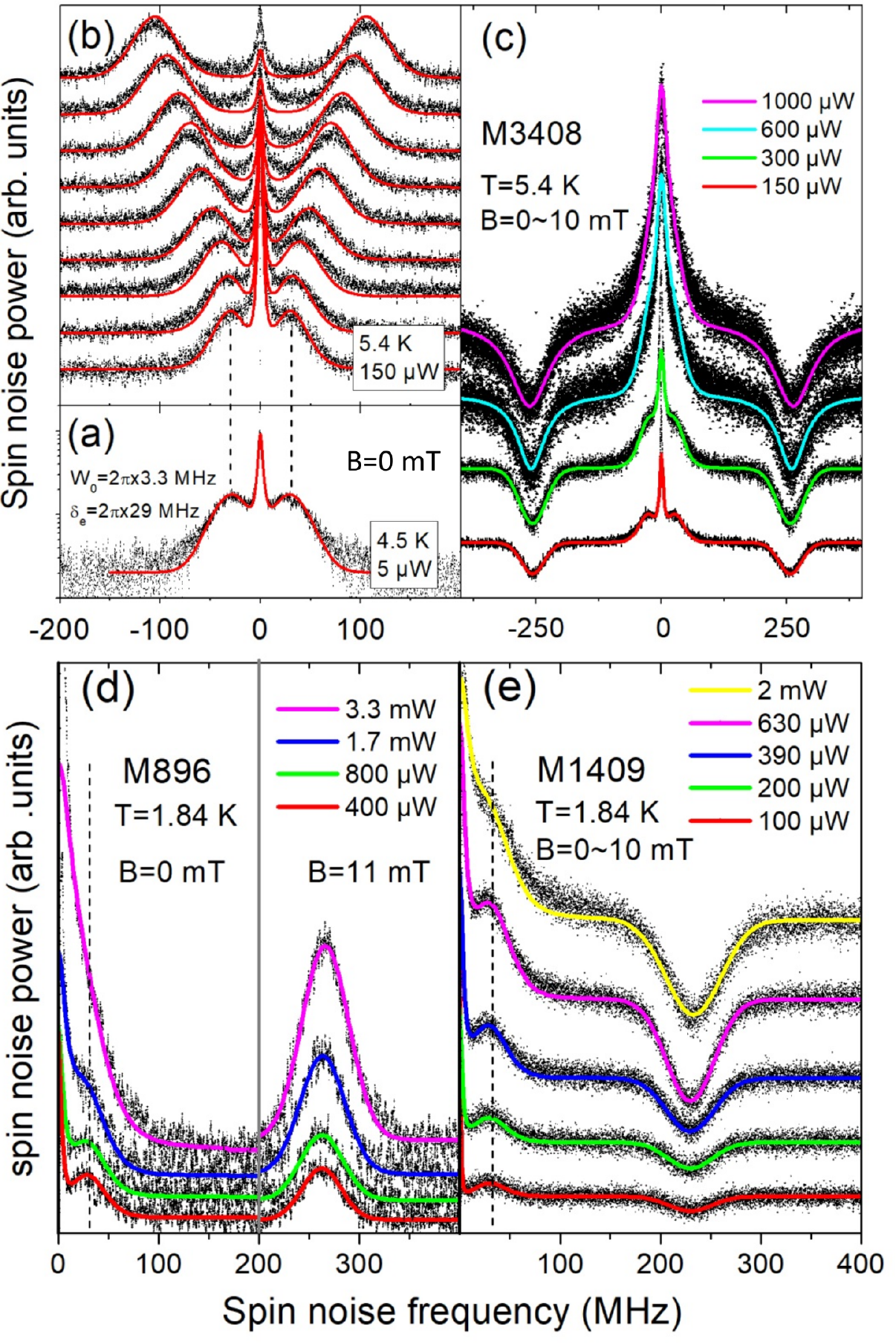}
  \caption{Spin noise spectra obtained from three different samples and different detection configurations: (a,b,c) M3408 and heterodyne detection; (d,e) M896, M1409 respectively and self-homodyne detection. (a) Zero magnetic field spin noise. (b) Series of noise spectra for increasing magnetic fields from 0 to 4~mT in steps of 0.5~mT. The red curves are fits of the data with Glazov's model \cite{Glazov2015a} ($W_0$ is the hopping rate, and $\delta_e$ represents the dispersion of the nuclear field distribution, see Sec. III B). (c,d,e) Series of noise spectra for different probe powers. The spectra in (c,e) are obtained by taking the difference between spectra measured at zero and non-zero magnetic field. The spectra in (d) are obtained either at zero magnetic field or in a fielf of 11~mT. In all panels the lines at zero magnetic field (close to zero frequency) are fitted with Glazov's model, while the lines in non-zero magnetic field (at frequencies above typically 200 MHz) are fitted with gaussian, excepted at 150 and 300~$\mu$W in (c) where the lines are fitted with lorentzian. In (b,c,d,e) the curves are vertically offset for clarity. }\label{Figure2}
  % (a) et(b) : M3408 n°46 à 54
  % (c) M3408 n°40 à 45
  % (d) M896 n°29 à 33
  % (e) M1409 n°25 à 28 +31 f=230MHz --> g=1.64
 \end{figure}

Figure~\ref{Figure2} shows  examples of spin noise spectra obtained from samples M3408, M896, and M1409. The satellite is the line centered at 29~MHz in zero magnetic field (Fig.~\ref{Figure2}) (position outlined by vertical dashed lines in (a,b,d,e)). The satellite is better seen at low probe powers and disappears at high powers (Fig.~\ref{Figure2}(c,d,e)). This behaviour is common to all samples in which the satellite has been detected. The frequency at the peak of the satellite corresponds to the Larmor precession in a magnetic field of about 1.2~mT. We checked, in particular in sample M896, that this effective field cannot be compensated by an external magnetic field. A magnetic field of $\sim1$~mT was applied in eight different high symmetry crystal directions perpendicular to the growth axis every 45°, and the same high frequency shift of the line was observed. This means that the electrons which contribute to the satellite precess in randomly oriented effective magnetic fields. Examples of the shift of the satellite with magnetic field for a given field direction is reported in Fig.~\ref{Figure2}(b).

As summarized in Table~\ref{table:sample}, the satellite can be detected from most of the studied samples, including insulating samples with density below the Mott transition (about $1.2\times 10^{17}$~cm$^{-3}$ in CdTe), samples close to the Mott transition, and even one metallic sample. In sample M1559 the satellite could be observed both in reflection and transmission geometry.

In sample M3406 the satellite could not be observed, but a relatively high probe power (2.6~mW) was required to detect the spin noise in this sample. In these conditions the spin noise spectrum in presence of magnetic field is lorentzian. This is not compatible with the observation of the satellite as will be shown later. Samples M3459 and M3460 were designed specifically for forward scattering measurements. The satellite has been observed only in sample M3460 but one cannot exclude that it can show up also in sample M3459 if appropriate conditions were found.

\begin{figure}[ht]
  \centering
  \includegraphics[width=8 cm]{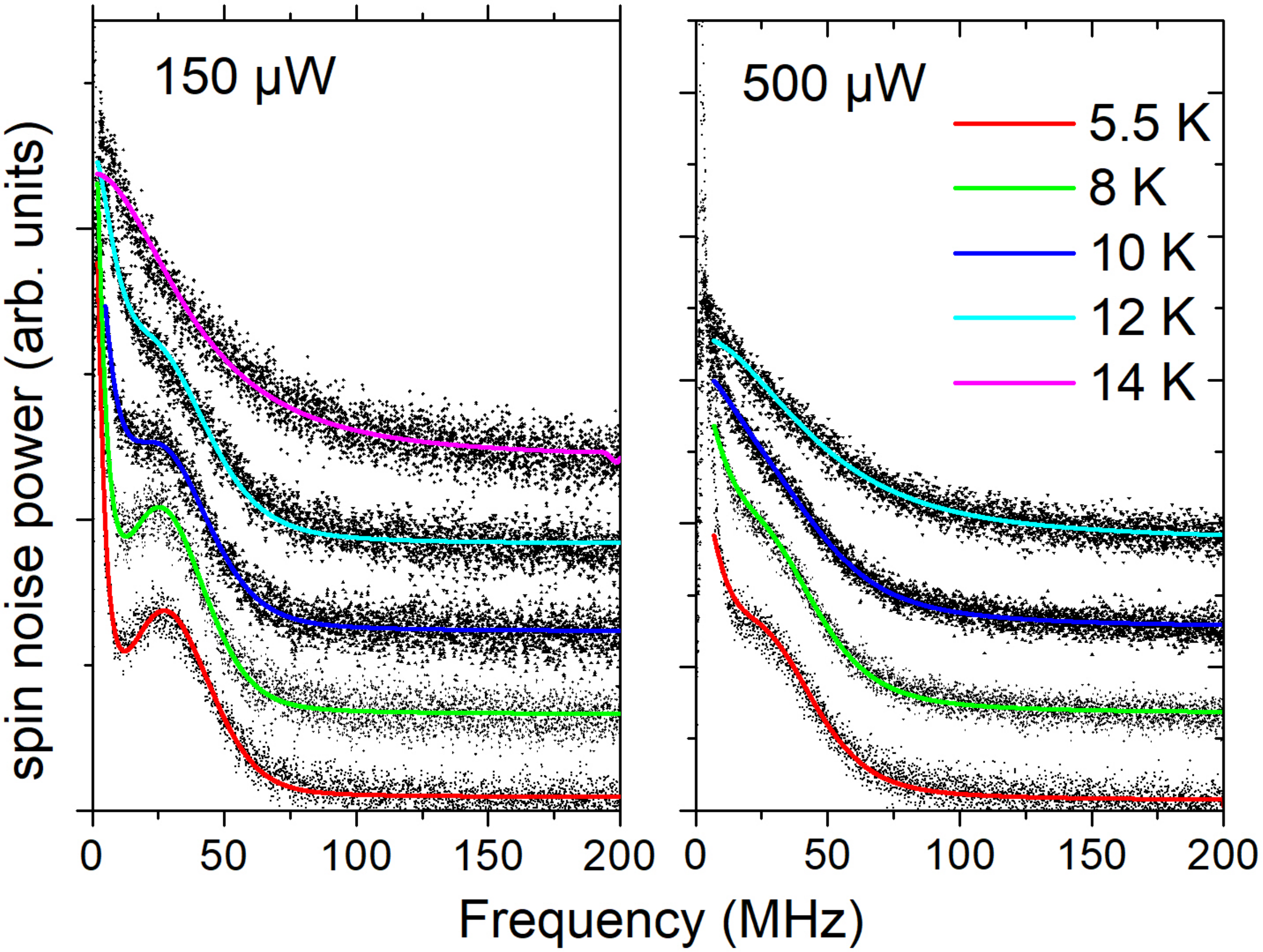}%from {serieB=0.jpg}
  \caption{Zero magnetic field spectra obtained from sample M3408 with the homodyne detection configuration. In this configuration the spin noise spectra cannot be reliably measured below 2-6 MHz due to low frequency noise, and the cutoff imposed by the detector response. }\label{Fig:serieB=0}
  %(C:\Users\Denis\ownCloud\n-CdTe\DATA\2019-02-07\07fevrierserieq)
\end{figure}

%\subsubsection{Temperature dependence}
Figure~\ref{Fig:serieB=0}(c,d,e) shows that the satellite disappears when the temperature is increased.  It also disappears with increasing power (Fig.~\ref{Figure2}(c)). This latter point results probably from laser heating.
The disappearance of the satellites at zero magnetic field is accompanied by a change of the lineshape of spin noise spectra under magnetic field, which evolve from gaussian to lorentzian and become broader as temperature or probe power increases (Fig.~\ref{Figure6}b). This transition from inhomogeneous to homogeneous broadening occurs when the correlation time of the effective random fields acting on the electrons becomes shorter than the Larmor period in the random field. In our case, this correlation time corresponds to the electron spin correlation time on a given donor site $\tau_c$, and the transition occurs for $\tau_c\delta_e<1$, or equivalently for $W_0>\delta_e$, where $W_0$ is the hopping rate of the electron spin, and $\delta_e$ the dispersion of the nuclear field distribution. A narrowing of the line should be expected due to motionnal averaging of the nuclear fields in contradiction with our observations. This probably means that another spin relaxation mechanism takes over in this case.

\subsection{Interpretation as spin precession in frozen nuclear spin fluctuations}\label{Sec:Overhauser}
A more detailed study of the satellite was performed with sample M3408: the spin noise was detected in backscattering geometry using the heterodyne detection configuration.

We interpret the satellite as being due to the spin precession of localized electrons in the quasi-static random nuclear field. This effect has already been observed in GaAs \cite{Berski2015a}, but only at very low doping $n_D\approx 1\times10^{14}\text{ cm}^{-1}$.

The peak centered at $\nu=0$ corresponds to the electron spin component along the nuclear field. Its width is determined by the longitudinal spin relaxation time (see Sec.~\ref{Sec:longitudinal}). The satellite corresponds to the electron spin precession in the random nuclear fields. In order to describe this characteristic two-peak structure we applied an electron spin noise model that takes into account the electron hopping between donors at a rate $W_0$ (the inverse of the electron correlation time $\tau_c$), the electron spin precession in random frozen nuclear fields determined by $\delta_e$, and the spin relaxation rate not related to hyperfine interaction and hopping $\nu_s$ \cite{Glazov2015a}. The agreement between experiment and theory is quite good, and noticeably the model reproduces very well the relative intensities of the central line and the satellite. The spin noise spectrum shown in Fig.~\ref{Figure2}(a) can accurately be fitted with $W_0=2\pi\times3.3$~MHz, $\nu_s=0$, and $\delta_e=2\pi\times29$~MHz. In these conditions ($W_0,\nu_s\ll\delta_e$) the two-peak structure can be approximated by the sum of a Maxwell distribution $S_M(\nu)=\frac{2(2\pi)^3A_M}{\sqrt{\pi}\delta_e^3}\nu^2\exp(-\frac{(2\pi\nu)^2}{\delta_e^2})$ and a lorentzian line $S_L(\nu)=2A_L\gamma_e/((2\pi\nu)^2+\gamma_e^2)$, with $\gamma_e=\nu_s+\frac{2}{3}W_0$. We also obtain a good agreement with the ratio $A_L/A_M=0.57$ close to the expected 1/2 theoretical value \cite{Glazov2015a}. However, the model taking into account the hopping between donors better reproduces the shape of the dip between the Lorentz and Maxwell lines. Note that the value of $\gamma_e$ is strongly constrained by the width of the central line, while the individual values of $\nu_s$ and $W_0$ are more loosely defined.

For an electron localized on a donor $\delta_e$ can be expressed as \cite{Merkulov2002}
\begin{equation}\label{delta_e}
  \delta_e=\frac{1}{\hbar}\sqrt{\frac{I(I+1)}{12\pi a_B^3}v_0\sum_\alpha A_\alpha^2\rho_\alpha},
\end{equation}
where $a_B$ is the donor Bohr radius, $v_0$ is the volume of the unit cell, $A_\alpha$ is the hyperfine constant for the isotope $\alpha$, $\rho_\alpha$ is the isotopic abundance. All nuclei bearing a spin in CdTe have the same spin $I=1/2$.
The hyperfine constant is given by
\begin{equation}\label{Ahyp}
  A_\alpha=\frac{2}{3}g_0\mu_0\mu_B\mu_\alpha |\psi_\alpha|^2/I,
\end{equation}
where $\mu_\alpha$ are the nuclear magnetic moments, and $|\psi_\alpha|^2$ are the electronic densities at the positions of the nuclei.
For the calculation of $\delta_e$ we take $a_B=5.2\text{ nm}$ and $v_0=45.16$~{\AA}$^3$. The stable isotopes of Cd and Te with non-zero spins are listed in Table \ref{table hyp} with their abundances and magnetic moments.
 The hyperfine constants $A_\alpha$, and the electronic densities $|\psi_\alpha|^2$, are poorly known in CdTe. To our knowledge, the only experimental determination of $|\psi_\text{Cd}|^2=6.5\times  10^{31}\text{ m}^{-3}$ was  obtained from measurements of the Knight shift in metallic n-doped CdTe by Look and Moore \cite{Look1972}, but the value of the electron $g$-factor at the Fermi energy ($g_\text{F}=-1.1$), that they use to deduce the electronic density from the Knight shift, is too small compared to theoretical expectations.
 For this reason Nakamura et al \cite{Nakamura1979} corrected this value  taking $g_\text{F}=-1.4$ obtained in a four-band k-p perturbation theory \cite{Lax1961}.
 They obtained $|\psi_\text{Cd}|^2=5.3\times  10^{31}\text{ m}^{-3}$. Since that time, more realistic k-p theory has shown that one must include the spin orbit interaction to account for the energy dependence of the electron $g$-factor in semiconductors with bulk inversion asymmetry \cite{Pfeffer2012}.
 With this more accurate theoretical estimate of the $g$-factor a slightly larger value $|\psi_\text{Cd}|^2=5.6\times 10^{31}\text{ m}^{-3}$ is obtained. This new value remains smaller than $|\psi_\text{Ga}|^2=5.8\times 10^{31}\text{ m}^{-3}$ in GaAs \cite{Paget1977}, and much smaller than $|\psi_\text{In}|^2=9.3\times 10^{31}\text{ m}^{-3}$ in InSb \cite{Gueron1964}, in disagreement with the expected chemical trend according to which the electronic density on the nucleus should increase with the atomic number \cite{Gueron1964}. One must keep in mind that the above determination of $|\psi_\text{Cd}|$ relies on a single measurement of the Knight shift, and a theoretical estimate of $g_\text{F}$ at a Fermi energy $E_\text{F}\simeq 40\text{ meV}$ \footnote{We note that there is a fair agreement between the calculated \cite{Pfeffer2012} and experimental \cite{Oestreich1996,Ito2009} temperature dependence of the electron $g$-factor in CdTe.}.
We could not find any experimental determination of $A_\text{Te}$ or $|\psi_\text{Te}|$ in the literature \footnote{The value given in Ref. \cite{Testelin2008} is deduced from the value of $A_\text{Cd}$ assuming $|\psi_\text{Te}|=|\psi_\text{Cd}|$}.
 If one keeps the up-dated $|\psi_\text{Cd}|=5.6\times 10^{31}\text{ m}^{-3}$ one must increase $r=|\psi_\text{anion}|^2/|\psi_\text{cation}|^2$ up to 2.5 in order to fit with $\delta_e=2\pi\times 29$~MHz. This value is well above $r=1.7$ generally accepted for GaAs and InSb. In III-V compounds the cation-anion bond is close to neutral, while in CdTe it is more ionic. This means that an electron at the bottom of the conduction band should sit preferentially on the cation, therefore one expects $r\leq1.7$.
 This leads us to admit that $|\psi_\text{Cd}|$ is still underestimated.  Hence, we treated $|\psi_\text{Cd}|^2$ as an adjustable parameter, but kept $r==1.7$ as for GaAs and InSb.
$\delta_e=2\pi\times29$~MHz is obtained for $|\psi_\text{Cd}|=7.6\times 10^{31}\text{ m}^{-3}$ in better agreement with the chemical trend. This value represents 72\% of the atomic value \footnote{Deduced from the hyperfine splitting of the Cd$^{+}$ ion \cite{Hamel1973}.}, similar to the reduction observed for different atoms in III-V and II-VI compounds. Table \ref{table hyp} gives the hyperfine constants deduced from the $\mid\psi_\alpha\mid^2$.

\medskip
\begin{table*}[ht]
  \centering
  \begin{tabular}{c c c c c}
  \hline
 $\alpha$ & $\mu_\alpha$ (in nuclear magneton) & $\rho_\alpha$ & $|\psi_\alpha(0)|^2$ (m$^{-3}$) & $A_\alpha$ $\mu$eV \\
 \hline
  $^{111}$Cd & -0.59 & 0.12 & $7.6\times 10^{31}$ & -44 \\
  \hline
  $^{113}$Cd & -0.62 & 0.12 & $7.6 \times 10^{31}$ & -46 \\
  \hline
   $^{123}$Te & -0.737 & 0.0089 & $12.9 \times 10^{31}$ & -93 \\
  \hline
   $^{125}$Te & -0.888 & 0.071 & $12.9 \times 10^{31}$ & -112 \\
  \hline
\end{tabular}
  \caption{Parameters used for the calculation of $\delta_e$. }\label{table hyp}
\end{table*}

As a by-product of our study focusing on the spin dynamics of the electrons in n-doped CdTe layers, we deduce the maximum nuclear field for fully aligned nuclear spins $B_\text{S}=I\sum_\alpha A_\alpha \rho_\alpha/g_e\mu_B\simeq 0.10\text{ T}$ ( with $g_e=-1.65$ for the electron $g$-factor \cite{Oestreich1996}). It is interesting to note that although $B_\text{S}$ is about 50 times larger in GaAs than in CdTe, $\delta_e$ has almost the same value \cite{Berski2015a}. This is a consequence of the combined effect of smaller Bohr radius, and larger electron $g$-factor in CdTe compared to GaAs.

\section{Spin diffusion studied by spatiotemporal SNS}\label{Sec:diffusion}

The spin diffusion coefficient is measured by $q$-selective homodyne detection. Mixing of the LO with spin-flip Raman scattered light at some angle allows to detect selectively the spin fluctuations characterized by a wavevector $q$, where $q$ is the difference between the incident (probe) and scattered (selected by the LO) wavevectors in the spin-flip scattering process \cite{Cronenberger2019}.

\begin{figure}[ht]
  \centering
  \includegraphics[width=8 cm]{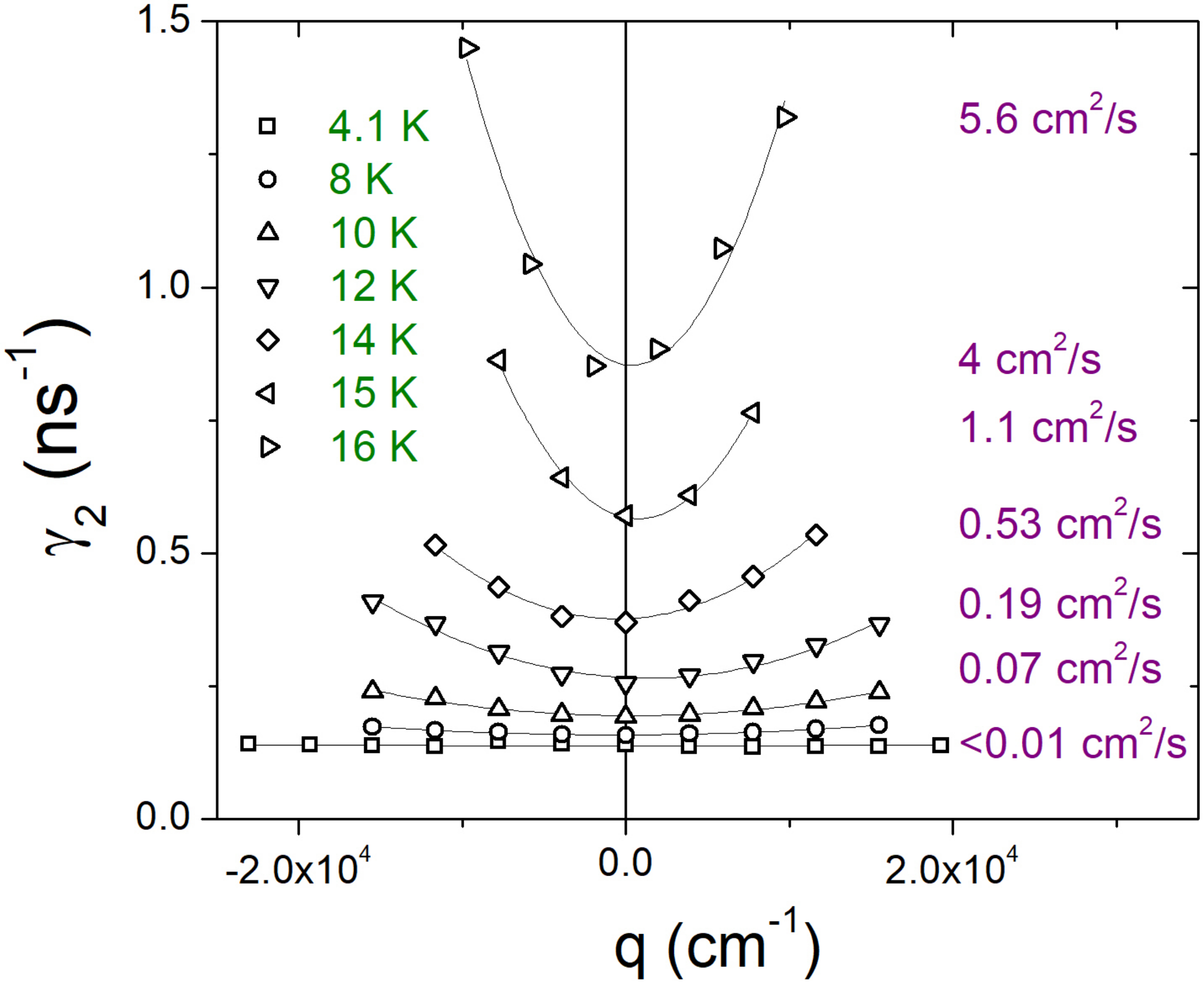}%from {diffusion.jpg}
  \caption{Broadening parameter $\gamma_2$ versus $q$ obtained from sample M3408 for a set of temperatures (open symbols), and quadratic fits (continuous lines) from which the spin relaxation rate $\gamma_s$ and spin diffusion coefficient $D_s$ are deduced (probe power 500 $\mu$W, B=22-24.4 mT). %From origin project M3408-homodynage29janv19.
  }\label{Fig:diffusion}
\end{figure}

\begin{figure}[ht]
  \centering
  \includegraphics[width=8 cm]{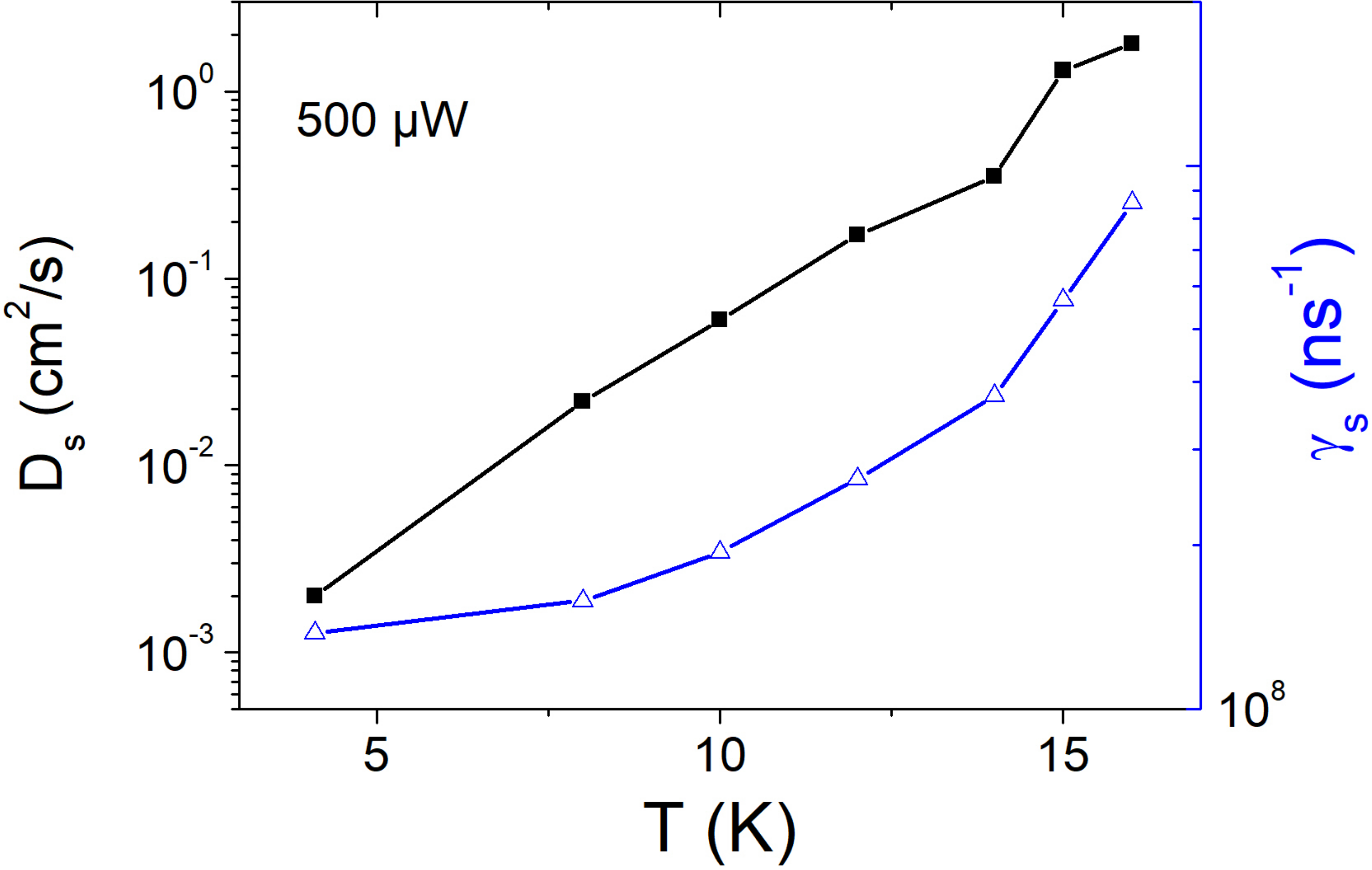}%from{D&gamma.jpg}
  \caption{$D_s$ and $\gamma_s$ deduced from the fits shown in Fig.~\ref{Fig:diffusion}}\label{Fig:D&gamma}
\end{figure}

Figure (\ref{Fig:diffusion}) shows the (transverse) spin relaxation rate $\gamma_2$ measured as a function of $q$ for different temperatures. One can see that $\gamma_2$ increases quadratically with $q$, as expected in presence of spin diffusion.
The broadening of the spin noise line in presence of spin diffusion is given by
\begin{equation}\label{spin_diff}
  \gamma_2(q)=\gamma_s+D_s q^2
\end{equation}
From quadratic fits of the data we obtain the spin diffusion coefficient $D_s$ and the spin relaxation rate $\gamma_s$, both shown in Fig.~(\ref{Fig:D&gamma}) and reported in Table~\ref{Table3}.

\medskip
\begin{table}[ht]
  \centering
  \begin{tabular}{c c c c c}
  \hline
 T (K) & P ($\mu$W) & $D_s$ (cm$^{2}$/s)  & $\gamma_s$ (ns$^{-1}$)& $W_0 (\mu\text{s}^{-1}$) \\
 \hline
  8 & 500 & 0.07 & 0.15& 44.7 \\
  10 & 500 & 0.19 & 0.18 & 94.4 \\
  12 & 500 & 0.53  & 0.25& 194.8 \\
  14 & 500 &1.1  & 0.38 & -\\
  15 & 500 &4  & 0.57 & - \\
  16 & 500 &5.6  & 0.85& - \\
  \hline
  8 & 400 & 0.01 & 0.13 & 6.2 \\
  10 & 400 &0.04 & 0.15 & 18.8 \\
  12 & 400 &0.075  & 0.17 & 37.7 \\
  \hline
  5.5 & 150 &0.05  & 0.16 & 12.5 \\
  12 & 150 &0.22  & 0.14 & 56.5 \\
  \hline
\end{tabular}
  \caption{Summary of the results obtained from sample M3408 for three series of measurements taken at different temperatures (first column), and different probe powers (second column). Third and fourth columns are the values of the spin diffusion $D_s$ and of the spin relaxation rate $\gamma_s$ measured by spatiotemporal SNS. The fifth column is the value of the hopping rate $W_0$ obtained from the fits of the two-peak structure at zero magnetic field. The data at 500~$\mu$W are obtained from the measurements shown in the right panel of Fig.~\ref{Fig:serieB=0} and in Fig.~\ref{Fig:diffusion}. }\label{Table3}
\end{table}

\section{Interpretation and discussion}\label{Sec:interpretation}
The spin of localized electrons can relax because of anisotropic exchange interaction between donors (AEI) \cite{Kavokin2001} or through the hyperfine interaction (HI) \cite{Dzhioev2002}. The spin relaxation is then driven by hopping  either of the spin alone by exchange between two occupied donors or hopping of the electron from an occupied to an unoccupied donor. We neglect the Dyakonov-Perel relaxation mechanism for delocalized electrons, because it is hardly compatible with the long spin correlation times we are dealing with.

The spin relaxation rate for the AEI is given by
\begin{equation}\label{AEI}
  \gamma_\text{AEI}=\frac{2}{3}\frac{\theta^2}{\tau_\text{hop}},
\end{equation}
where $\tau_\text{hop}$ is the hopping time, and
\begin{equation}\label{theta}
  \theta=\frac{\gamma_D}{E_ba_B^3}\left(0.323+0.436r+0.014r^2\right)
\end{equation}
 is the rotation angle of the spin due to hopping between donors \cite{Dzhioev2002,Kavokin2001}. $\gamma_D$ is the Dresselhaus coefficient, $E_b$ the donor binding energy, $a_B$ is the Bohr radius, and $r=r_c/a_B$. $r_c$ is the characteristic distance between interacting donors and is given by $r_c=\beta n_d^{-1/3}$, with $0.54\le\beta\le 0.8$.

The spin relaxation rate due to hyperfine interaction is given by
\begin{equation}\label{HI}
  \gamma_\text{HI}=\frac{2}{3}\langle \omega_N^2\rangle \tau_c=\frac{2}{3}\delta_e^2 \tau_c
\end{equation}
in the motionnal narrowing regime where the spin correlation time $\tau_c \ll\delta_e^{-1}$. In the opposite case the line is inhomogeneously broadened by the distribution of nuclear fields and the spin relaxation rate, defined as the FWHM divided by $\pi$, is $\gamma_s=\sqrt{\log(2)}\delta_e$. In the following we will assume that the correlation time is only limited by hopping of electrons on donor sites, so that $\tau_c=\tau_\text{hop}$.
Hence, the diffusion constant is given by $D_s=r_c^2/6\tau_c$.

\begin{figure*}[ht]
  \centering
  \includegraphics[width=17 cm]{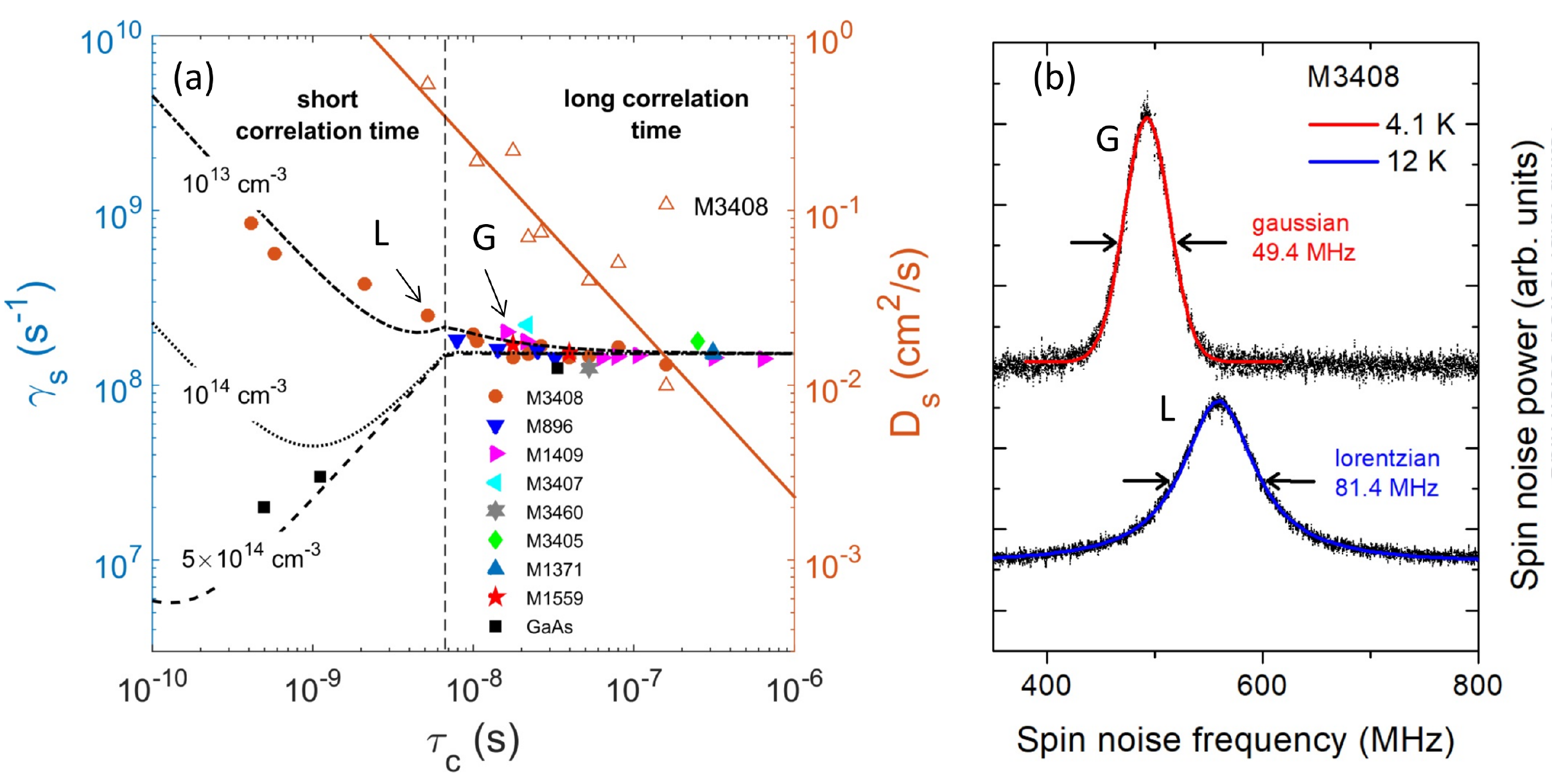}
  \caption{(a) Comparison between measured (symbols) and calculated (lines) $\gamma_s$ and $D_s$ versus $\tau_c$.   The black curves are the calculated spin relaxation rates for different donor concentrations in CdTe (dotted, dashed-dotted), and in GaAs (dashed). The solid brown line is the calculated spin diffusion constant $D_s$. The filled, colored symbols are the measured spin relaxation rates for all samples where the satellite was observed. The open triangles are the measured spin diffusion coefficients for sample M3408. The black filled squares are spin relaxation rates measured in GaAs and taken from literature \cite{Berski2015a,Dzhioev2002}. (b) Examples of spin noise spectra measured in the regime of long and short correlation times (corresponding to the data points marked as G, and L respectively in (a). }\label{Figure6}
  %figure generate by matlab script ExchangeInteractionBetweenDonors
\end{figure*}

In Figure~\ref{Figure6} we compare the measured $\gamma_s$ and $D_s$ versus $\tau_c$ with the calculated ones.
The data for sample M3408 are taken from Table~\ref{Table3} with $\tau_c=W_0^{-1}$.
The measured $D_s$ follows reasonably the expected $\tau_c^{-1}$ dependence for the data where $W_0$ is available (in the long correlation time regime). For the missing values of $W_0$, we assume that this dependence is still satisfied in the short correlation time regime. This allows to deduce the value of $\tau_c$ from the measured $D_s$, and plot the three remaining data points of Table~\ref{Table3}.

We also measured $\gamma_s$ and $D_s$ from samples M3459, M3460, and M3416. In some conditions we find $\gamma_s\le1\times10^8\text{ s}^{-1}$, i.e. less than the relaxation rate due to hyperfine fields in the long correlation time regime. This may indicate that the motionnal narrowing regime is reached. However, we could not detect the satellites in samples M3459 and M3416, therefore it was not possible to estimate $W_0$ and $\tau_c$ in these samples. In sample M3460 the satellite was evidenced but joint measurements of the spin diffusion coefficient are lacking.
For most of the samples where we did not measure $D_s$, but could observe the satellites, data in the long correlation time regime are shown.

The dotted curve represents $\gamma_s$ calculated with the parameters of CdTe ($E_b=10\text{ meV}$, $a_B=5.2 \text{ nm}$, $\delta_e=2\pi\times29\text{ MHz}$) with $n_d=10^{14}\text{ cm}^{-3}$ and $\gamma_D=12 \mu\text{eV\AA}^3$ in the range of published values \cite{Cardona1988,Petrovic2012,Wojtowicz2018}.
For this doping level one can distinguish the three different spin relaxation regimes. For $\tau_c<10^{-9} \text{ s}^{-1}$ the relaxation is dominated by the AEI, thus $\gamma_s$ and $\tau_c$ are inversely proportional. For $\tau_c>10^{-9} \text{ s}^{-1}$ the relaxation is dominated by the HI, in the motionnal narrowing regime up to $5\times 10^{-9} \text{ s}^{-1}$. For $\tau_c>5\times 10^{-9} \text{ s}^{-1}$  one enters in the regime of pure spin dephasing where the broadening is constant.

Increasing the doping level shifts the transition between the AEI and the HI relaxation regimes to smaller $\tau_c$. The dashed curve illustrates this case. It has been calculated using the parameters of GaAs ($\gamma_D=19 \mu\text{eV\AA}^3$, $E_b=5\text{ meV}$, $a_B=10 \text{ nm}$, $\delta_e=2\pi\times29\text{ MHz}$) to allow a comparison with data for GaAs taken from literature. Note however that the value of $\delta_e$ is practically the same in CdTe and GaAs, therefore the calculated curves are identical in the HI regime. In GaAs, the regime of dynamic averaging of the random nuclear fields is evidenced by the decrease of $\gamma_s$ when $\tau_c$ decreases, whereas for CdTe $\gamma_s$ increases as $\tau_c$ decreases. Here shorter $\tau_c$ are reached on the same sample by increasing the temperature, while in GaAs the measurements were all performed at low temperatures and $\tau_c$ is mainly correlated to the doping level. It is plausible that in this case some other spin relaxation mechanism comes into play. Nevertheless we try to fit our data keeping only the two spin relaxation mechanisms described above. It is remarkable that for $n_d=10^{13}\text{ cm}^{-3}$ one obtains a good agreement between theory and experiment, simultaneously for $\gamma_s$ and $D_s$ without additional fitting parameters.

To date, the most systematic study of electron spin relaxation in n-doped CdTe has been performed by time-resolved Kerr rotation experiments \cite{Sprinzl2010}. The longest spin relaxation time was found to be $\simeq2.5$~ns for $n_d\simeq5\times10^{16}\text{ cm}^{-3}$. This relaxation time sharply decreases at lower concentrations to reach about 40~ps for $n_d<10^{16}\text{ cm}^{-3}$. This behaviour seems hardly compatible with the now well understood spin relaxation mechanisms established for GaAs. In fact, due to the quite similar local nuclear fields seen by electrons localized on donors, and the similar values of the Dresselhaus constant in GaAs and CdTe, one could expect similar spin relaxation properties in both materials. Our results differ considerably from those obtained in \cite{Sprinzl2010}, since we measured longer spin relaxation times $\sim 6-7$~ns for a broad range of doping levels. In most cases, this relaxation can be explained by HI only.
%\medskip
\section{Longitudinal spin relaxation}\label{Sec:longitudinal}
Until now we have considered the transverse spin relaxation rate $\gamma_s$ related to the satellite. We now turn to the longitudinal spin relaxation related to the line centered at zero spin noise frequency (see Fig.~\ref{Figure2}). The longitudinal relaxation rate $\gamma_1$ is limited by $W_0$ and by spin relaxation mechanisms other than the interaction with nuclei.

\begin{figure}[ht]
  \centering
  \includegraphics[width=8 cm]{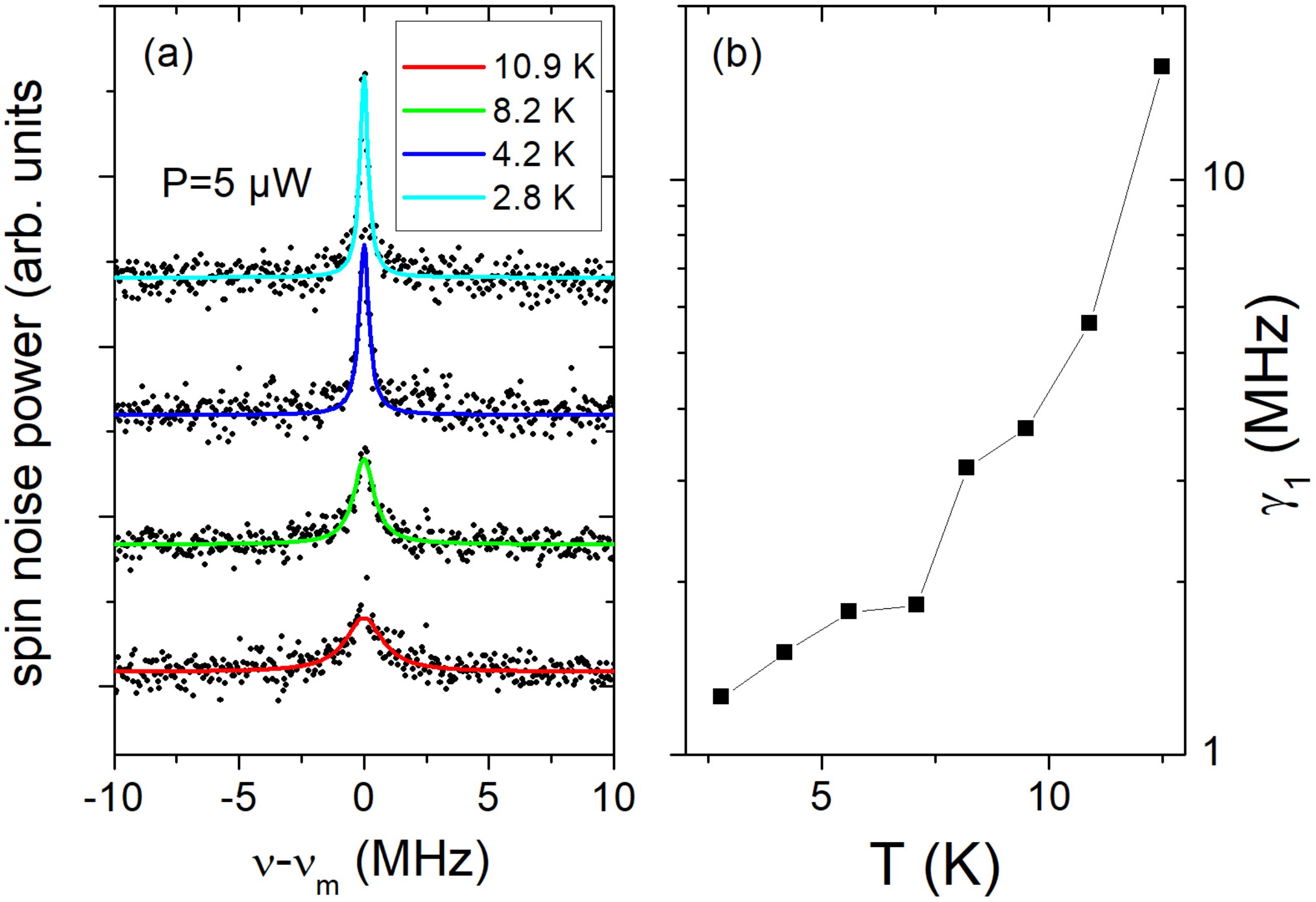}
  \caption{(a) Measured spectra together with lorentzian fits. (b) Longitudinal spin relaxation rates.}\label{Figure7}
  %data : 2018/03/26      M3408-063 -> 070
\end{figure}

We find that the linewidth of the central line strongly decreases as the laser power is reduced down to 5~$\mu\text{W}$. The linewidth also decreases rapidly with decreasing temperature (Fig.~\ref{Figure7}), and at the lowest temperature we obtained the spin relaxation time $\tau_s=\gamma_1^{-1}=0.8\text{ }\mu\text{s}$. This is not an absolute limit since longer relaxation times up to $\sim1$~ms have been measured in CdTe at lower doping density, and lower temperatures \cite{Linpeng2016}. Note that at very low powers the satellite lines become hardly visible because they are much broader than the zero-frequency line.
\section{Conclusion}
We measured spin noise spectra for n-doped CdTe layers with electron densities varying across the Mott transition value and detected rather systematically a satellite line in the zero magnetic field spectra due to the local nuclear fields. This allows us to show that for fully polarized nuclei the Overhauser field can reach 0.10~Tesla in CdTe. The existence of the satellite also implies long spin correlation times $\tau_c>6$~ns, even for relatively high doping levels, which is quite surprising and not understood. In addition we used the spatiotemporal SNS technique, which allows to determine simultaneously $\gamma_s$ and $D_s$. Probably a complete understanding of electron spin relaxation mechanisms in CdTe will require a joint study of spin relaxation and transport properties similar to what has been done for GaAs \cite{Lonnemann2017}.

%\bibliographystyle{apsrev4-1}
%\bibliography{library}


\end{document}